\documentclass[twocolumn,aps,prl] {revtex4}
\usepackage{graphicx,subfigure}

\begin{document}

\title{Velocity dependence of friction of confined polymers}

\author{I.M. Sivebaek}
\affiliation{IFF, FZ-J\"ulich, 52425 J\"ulich, Germany}
\affiliation{Novo Nordisk A/S, Research and Development, DK-3400 Hillerod, Denmark}
\affiliation{Mech. Eng. Dept., Technical University of Denmark, DK-2800 Lyngby, Denmark}

\author{V.N. Samoilov}
\affiliation{IFF, FZ-J\"ulich, 52425 J\"ulich, Germany}
\affiliation{Physics Faculty, Moscow State University, 117234 Moscow, Russia}

\author{B.N.J. Persson}
\affiliation{IFF, FZ-J\"ulich, 52425 J\"ulich, Germany}

\begin{abstract}

We present molecular dynamics friction calculations for confined hydrocarbon solids with
molecular lengths from 20 to 1400 carbon atoms. Two cases are considered: (a) polymer
sliding against a hard substrate, and (b) polymer sliding on polymer. We discuss the
velocity dependence of the frictional shear stress for both cases.
In our simulations, the polymer films are very thin ($\sim 3 \ {\rm nm}$), and the solid walls
are connected to a thermostat at a short distance from the polymer slab. Under these
circumstances we find that frictional heating effects are not important, and the effective temperature
in the polymer film is always close to the thermostat temperature.

In the first setup (a), for hydrocarbons with molecular lengths from $60$ to $1400$ carbon atoms,
the shear stresses are nearly independent of molecular length, but
for the shortest hydrocarbon C$_{20}$H$_{42}$ the frictional shear stress is lower.
In all cases the frictional shear stress increases monotonically with the sliding velocity.

For polymer sliding on polymer [case (b)]
the friction is much larger, and the velocity dependence is more complex.
For hydrocarbons with molecular lengths from 60 to 140 C-atoms, the number of monolayers of
lubricant increases (abruptly) with increasing sliding
velocity (from 6 to 7 layers), leading to a decrease of the friction.
Before and after the layering transition, the frictional shear stresses are
nearly proportional to the logarithm of sliding velocity. For the longest
hydrocarbon (1400 C-atoms) the friction shows no dependence on the sliding velocity,
and for the shortest hydrocarbon (20 C-atoms) the frictional shear stress increases
nearly linearly with the sliding velocity.

\end{abstract}
\maketitle


\subsection{1. Introduction}

Friction between solids is a very important phenomenon in
biology and technology \cite{P0} and it is very common in nature.
Static friction always involves the coexistence of different
metastable configurations at microscopic level.
When one surface slides on the other at low speed, first there is a
loading phase during which the actual configuration stores elastic
energy. Then, when the stored energy is large enough, an instability
arises \cite{Aubry,SSC,Nozier}: the system jumps abruptly to
another configuration and releases
elastic energy into irregular heat motion. The exact way of how the
energy is dissipated usually does not influence the sliding friction force,
provided that the dissipation is fast enough to happen before the next
sliding event.

There are many possible origins of elastic instabilities, e.g., they may
involve individual molecules or, more likely, groups of molecules or
``patches'' at the interface, which have been named stress
domains \cite{P11,P2,Caroli,Caroli1}.
Since the local rearrangements usually occur at different times in an incoherent
manner, at the macroscopic scale the sliding motion may appear smooth without
stick-slip oscillations. However, this
is always a result of self-averaging, and at the atomistic level stick-slip
motion will almost always occur (except for incommensurate systems with weak
interactions). Moreover, at least at zero temperature,
the friction force does not vanish in the limit of
sliding speed $v\to 0$, but it tends to some finite value which depends
on the average energy stored during the loading events and the atomic slip distance.

A logarithmic velocity
dependence of the frictional shear stress was observed in many experiments but usually for
sliding at low velocities (up to $\approx 20\ {\rm \mu m/s}$) \cite{riedognecco2004}.
Still, in some experiments the logarithmic velocity dependence of the frictional shear stress
was observed also for higher velocities (up to 1 mm/s) \cite{tocha2005}.
Comparing experimental results with the existing theoretical
models\cite{SangDubeGrant2008,P11,riedognecco2004} shows that the logarithmic
dependence could be well described in the models accounting for thermal activation effects.
Thus, thermal fluctuations may induce jump of atoms (or rather group of atoms)
at the sliding interface from one equilibrium position to the next
one along the reaction path. The resulting stress-aided thermally activated effect leads to a logarithmic
increase of friction with the velocity at low velocities. Thermal activation
is more efficient at low velocities, where the system spends long time in each potential well and, consequently,
the probability to thermally activate the processes of atoms hopping is higher.

When a polymer (or a long-chain alkane) is sheared between two surfaces the shear stress often does not
depend linearly function of the logarithm of the sliding velocity. This has been established both
experimentally \cite {bureau2006, drummond2002, qian2003} and in simulations \cite {subbotin1997}.
This observation may be due to the interdiffusion of chain segments between the
polymer layers by the long molecules. Thus, the formation of ``bridges'' is presumably affected by a change in the
sliding velocity and in certain regimes it is possible to have decreasing shear stress with
increasing sliding velocity due to the disappearing of bridges
across the sliding interface\cite {subbotin1997,qian2003}. This picture
was first suggested in the context of rubber friction by Schallamach\cite{Schallamach}.

In this paper we present molecular dynamics friction calculations for
confined hydrocarbon solids with molecular lengths from 20 to 1400 carbon atoms. Two cases
are considered: (a) polymer sliding against a hard substrate, and (b) polymer sliding on
polymer. We discuss the velocity dependence of the frictional shear stress for both cases.
We compare results obtained at room temperature and very low temperature (approaching 0 K).
In the latter calculations no thermal activation can occur.

In our simulations, the polymer films are very thin ($\sim 3 \ {\rm nm}$), and the solid walls
are connected to a thermostat at a short distance from the polymer slab. Under these
circumstances we find that frictional heating effects are not important, and the effective temperature
in the polymer film is always close to the thermostat temperature. In most practical
situations the temperature is not fixed at planes close to the interface and
during sliding at high enough velocities for a long enough time, the local
temperature at the sliding interface may be so high as to locally melt the polymer surfaces.
The physical processes occurring at the sliding interface in these cases may
be a combination of the effects studied in this paper and the influence of the increased temperature.

\subsection{2. The model}

In this paper we present computer simulation results about the
frictional behavior of linear hydrocarbons under applied pressure.
Our model is similar to those described in Refs. \cite{perbal2000,pervzn2002,SSP2003,SivebaekEPJE2008},
but we review its main features here. We consider a block and a substrate with
atomically flat surfaces separated by a polymer slab.
Two cases are considered: (a) polymer sliding against a hard substrate which we will
denote as ``metal'' for simplicity
(the metal-polymer case), and (b) polymer sliding on polymer (the polymer-polymer case).

The solid walls are treated as single layers of ``atoms'' bound
to rigid flat surfaces by springs corresponding to the
long-range elastic properties of $50 \ {\rm \AA}$
thick solid slabs.
For the case of sliding of polymer on ``metal'', all molecules are adsorbed on the
block surface only due to different parameters of interaction of alkane
molecules with the walls, whereas for the case of sliding of polymer on
polymer about half of the molecules adsorbed on the block surface and
half on the substrate surface. The block with adsorbed polymer slab was put into contact with
the substrate surface in the first case and two solids with adsorbed polymer slabs were put into
contact in the second case. When the temperature was equal to the thermostat temperature (usually 300 K) everywhere
we started to move the upper block surface.
We also conducted calculations for the temperature of solid
walls equal to 0 K in order to compare the sliding friction behavior at 300 K with that in
the absence of thermal activation at T = 0 K. The temperature was also varied from 300 K
to 550 K to study the effect of melting on the shear stress.

\begin{figure}
  \includegraphics[width=0.3\textwidth, angle=-90]{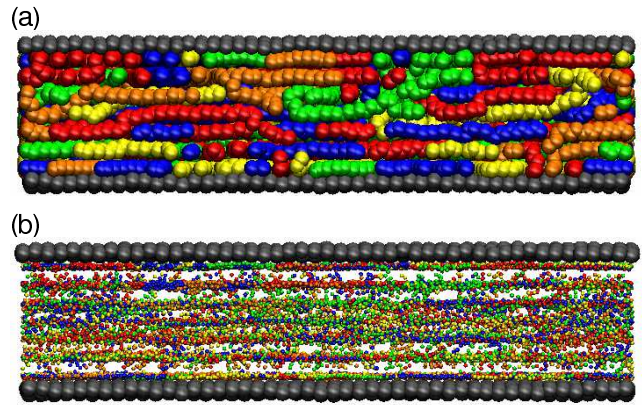}
  \caption{ \label{C100_p_plain_points}
Snapshot picture of the  ${\rm C}_{100}{\rm H}_{202}$ polymer slab
at the sliding velocity $v= 10 \ {\rm m/s}$ and
background temperature $T= 300 \ {\rm K}$. (a) The molecules are (arbitrarily) colored in order to
better observe the shear alignment of the chains. (b) The same as in (a) but with atoms presented
as points in order to observe layering in the system. Seven monolayers of molecules are clearly seen.
}
\end{figure}


Linear alkanes C$_n$H$_{2n+2}$ (with $n$ ranging from 20 to 1400) were used as
``lubricant'' in the present calculations.
The CH$_{2}$/CH$_{3}$ beads are treated
in the united atom representation \cite{jorgensen1984x1,dysthe2000x1}.
The Lennard-Jones potential
was used to model the interaction between beads of different chains
\begin{equation}
 \label{potential}
 U(r)=4 \epsilon_0 \left [ \left ({r_0\over r}\right )^{12}-\alpha
 \left ({r_0 \over r}\right )^{6} \right],
\end{equation}
and the same potential with modified parameters
$(\epsilon_1,r_1)$
was used for the interaction of each bead with the substrate and block atoms.
The parameters were $\epsilon_0 = 5.12 \ {\rm meV}$ for both the
interior and the end beads, and $r_0 = 3.905 \ {\rm \AA}$ and $\alpha=1$ in all cases.
For the interactions within the C$_n$H$_{2n+2}$ molecules we used the standard OPLS
model \cite{jorgensen1984x1,dysthe2000x1}, including flexible bonds, bond
bending and torsion interaction, which results in bulk properties in good agreement
with experimental data.

For polymer sliding on polymer we need the polymer-metal bond to be so strong that no
slip occurs at these interfaces. This is the case with $r_1 = 3.28 \ {\rm \AA}$,
$\epsilon_1 = 40 \ {\rm meV}$ and $\alpha=3$. We also did some simulations with
$\alpha=2$, but in this case some slip was observed at the polymer-metal interface.
For sliding of polymer on ``metal'' we used the same parameters as above for
the polymer-block interaction ({\em strong adsorbates interaction}) but with $\alpha=1$.
For the
polymer-substrate interaction we used
$\alpha=1$ and $\epsilon_1 = 10 \
{\rm meV}$ ({\em weak adsorbate interaction}).

The choice of higher values of $\epsilon_1$ compared to $\epsilon_0$ reflects the
stronger (van der Waals) interaction between the beads and the
``metal'' surfaces than between the bead
units of different lubricant molecules (this stronger interaction results from the
higher electron density in the metals).
The lattice spacings of the block and of the substrate are $a=b=2.6~\mbox{\AA}$.

We used linear alkane molecules with the number of carbon atoms 20, 60, 100, 140
and 1400 as lubricant. The number of C$_{100}$H$_{202}$ molecules was equal to 200.
The number of C$_{20}$H$_{42}$ molecules was equal to 1000. This gave from 6 to 8
monolayers of lubricant molecules between the solid surfaces.
The (nominal) squeezing pressure $p_0$ was usually $10 \ {\rm MPa}$.

As an illustration, in Fig.~\ref{C100_p_plain_points} we show the contact between a flat
elastic block (top) and a flat elastic substrate (bottom). The polymer slab ($\sim 30 \
{\rm \AA}$ thick) is in between them. Only the interfacial block and substrate atoms and
polymer atoms are shown.




\subsection{3. Polymer on ``metal''}

\begin{figure}
  \includegraphics[width=0.48\textwidth]
{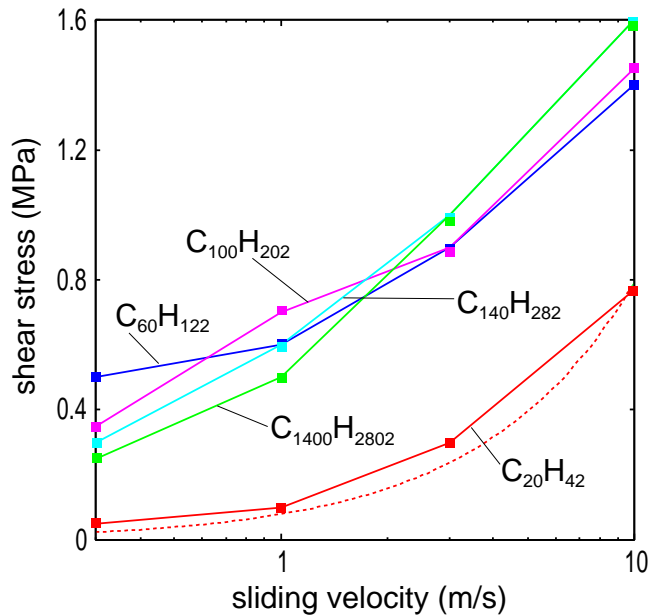}
  \caption{ \label{velocityCXXm10XX}
The dependence of the shear stress on the sliding velocity for
the polymers from C$_{20}$H$_{42}$ to C$_{1400}$H$_{2802}$ sliding on ``metal'' at the normal
pressure $10 \ {\rm MPa}$. The temperature $300 \ {\rm K}$. The dotted line is a linear fit to
the C$_{20}$H$_{42}$ curve.
}
\end{figure}

In Fig.~\ref{velocityCXXm10XX} we show the dependence of the shear stress on the
sliding velocity for the
polymer slabs sliding on ``metal'' at the applied pressure $p=10 \ {\rm MPa}$.
For C$_{20}$H$_{42}$ the lubricant behaves liquid-like
with the shear stress nearly proportional to the sliding velocity (see dashed curve in
Fig. \ref{velocityCXXm10XX}).
In particular, as the sliding velocity approaches zero the shear stress for C$_{20}$H$_{42}$
becomes very small, reflecting the fact that this polymer is close to the liquid state:
the C$_{20}$H$_{42}$ (eicosane) system is at
$27^\circ {\rm C}$ (the temperature of the thermostat),
which is very close to the melting point ($\approx 37^\circ {\rm C}$) of this polymer.
We also performed simulations at the (thermostat) temperatures $200 \ {\rm K}$
and $0 \ {\rm K}$, and in these cases the friction is considerably higher ($\sim 9.6$ times
higher at the temperature $0 \ {\rm K}$ for the sliding velocity $v=0.3 \ {\rm m/s}$).

For longer-chain hydrocarbon molecules we found that the polymer films behave
solid-like at $300\ {\rm K}$, with non-vanishing kinetic friction as the sliding velocity
approaches zero, and that the shear stress depends non-linearly on the
sliding velocity. This is due to the higher melting point of the longer-chain polymer
systems. For the C$_{100}$H$_{202}$ system the frictional shear stress
is more than twice higher at $T=0 \ {\rm K}$ than for $T=300 \ {\rm K}$ for the sliding
velocity $v=0.3 \ {\rm m/s}$.
This can be attributed to the absence of thermal activation at $T=0 \ {\rm K}$.
Atomistic stick-slip events occur at the sliding interface
between the polymer and the substrate.
The period of oscillations of the shear stress as a
function of $x$-coordinate of the block is exactly
$2.6 \ {\rm \AA}$ which is one lattice unit (of the substrate) in the sliding direction.
At low sliding velocities the shear stress
increases greatly (much more than for the higher sliding velocities) when the temperature is decreased
from $300 \ {\rm K}$ to $0 \ {\rm K}$. This is due to the fact
that the probability to activate the processes of atoms hopping is higher
at low velocities due to the longer time the system spends in each local potential well along the reaction coordinate.


We emphasize the importance of the temperature (or thermal fluctuations)
on the process of ``going over the barrier''. Thus at zero temperature, the external applied tangential
force (or stress) alone pulls the system over the lateral pinning barriers, and this happens
everywhere simultaneously.
At high sliding velocities thermal
effect should be rather unimportant.
However, for small
sliding velocities, thermal fluctuations will be very important. In this case
slip will not occur everywhere simultaneously, but
small nanometer-sized interfacial
regions of linear size $D$ will be individually pinned and perform stress-aided thermally induced
jump from one pinned
state to another (local interfacial rearrangement processes). (Note that thermal effects can only
become important for small (nanometer-sized $D$) regions, since
simultaneous going-over-the-barrier everywhere
requires infinitely large energy for an infinite system, except, perhaps for an incommensurate interface.)
This process has been studied in detail both
theoretically \cite{PerssonPRB,Briscoe,Schallamach,PV} and experimentally \cite{Caroli1,Drummond1,Baum}.

\begin{figure}
  \includegraphics[width=0.48\textwidth]{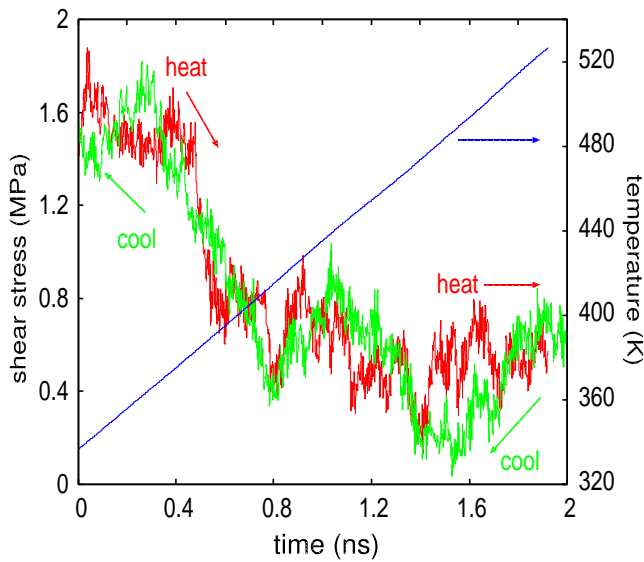}
  \caption{ \label{temperatureRampC140m1010}
The shear stress for C$_{140}$H$_{282}$ polymer sliding on metal at the sliding velocity
$v=10 \ {\rm m/s}$ and the normal pressure $p=10 \ {\rm MPa}$ when ramping (step-like
changing) the temperature $T$ from $300 \ {\rm K}$ to $550 \ {\rm K}$ (heating) and from
$550 \ {\rm K}$ to $300 \ {\rm K}$ (cooling).
}
\end{figure}

We now study the frictional behavior of the hydrocarbon films C$_{100}$H$_{202}$ and C$_{140}$H$_{282}$,
when the temperature is increased above the melting temperatures.
In Fig.~\ref{temperatureRampC140m1010} we show the shear stress for the C$_{140}$H$_{282}$
polymer film (at the sliding velocity $v=10 \ {\rm m/s}$ and the normal pressure
$p=10 \ {\rm MPa}$) when increasing the
thermostat temperature $T$ by steps equal to $50^\circ {\rm C}$ from $27^\circ
{\rm C}$ to $277^\circ {\rm C}$, and then decreasing it back to $27^\circ{\rm C}$.
During heating the shear stress decreases abruptly when the temperature is raised above
the melting temperature $T_{\rm m}=110^\circ{\rm C}$ for C$_{140}$H$_{282}$ polymer.
The frictional shear stress for the film in the liquid-like state is $\sim 3$ times lower
than for the solid film just below melting.

\begin{figure}
  \subfigure
  {
  \includegraphics[width=0.48\textwidth]{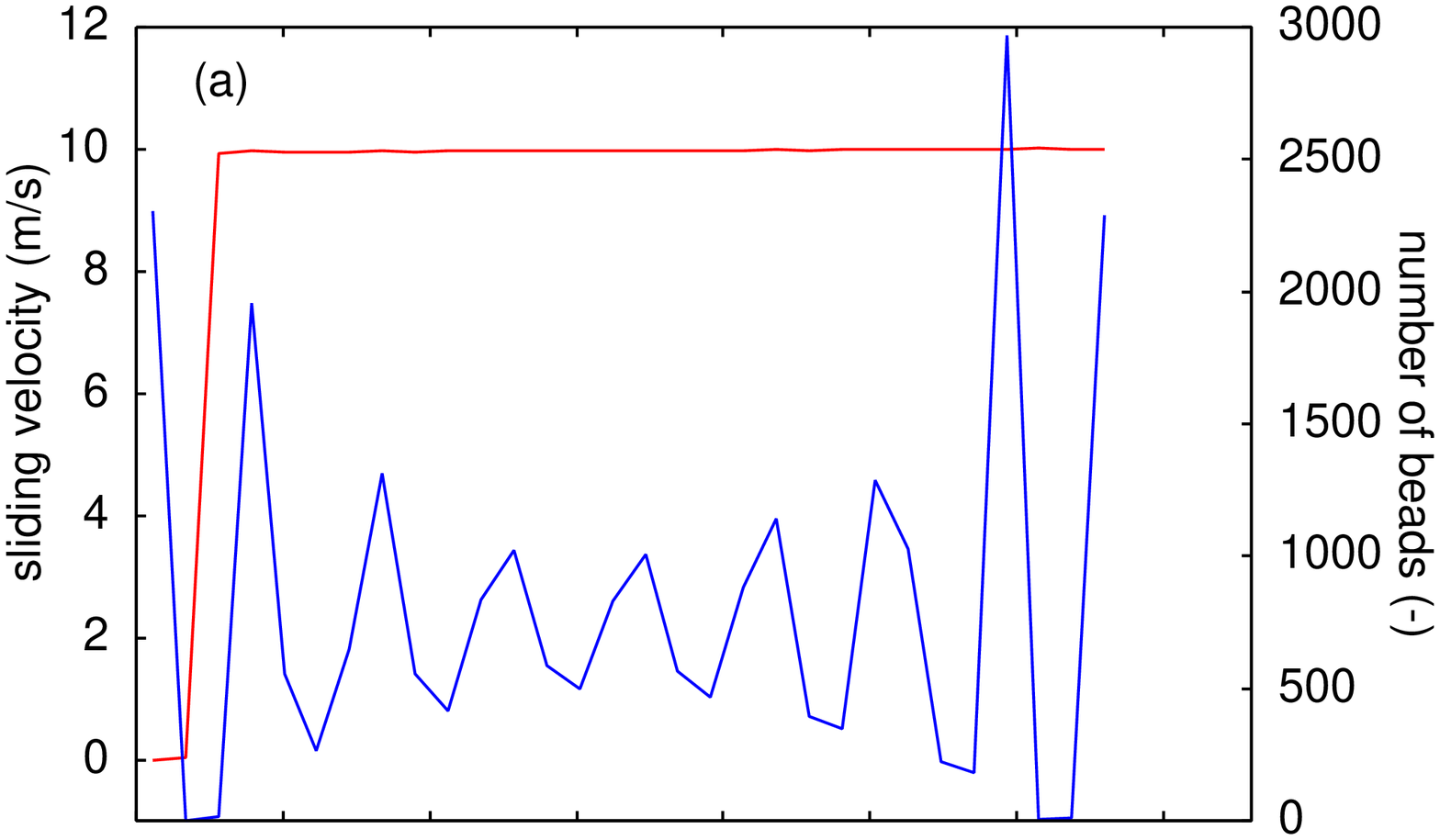}
  \label{velx_zposC100m1010_700_300_K}
  }
  \subfigure
  {
  \includegraphics[width=0.48\textwidth]{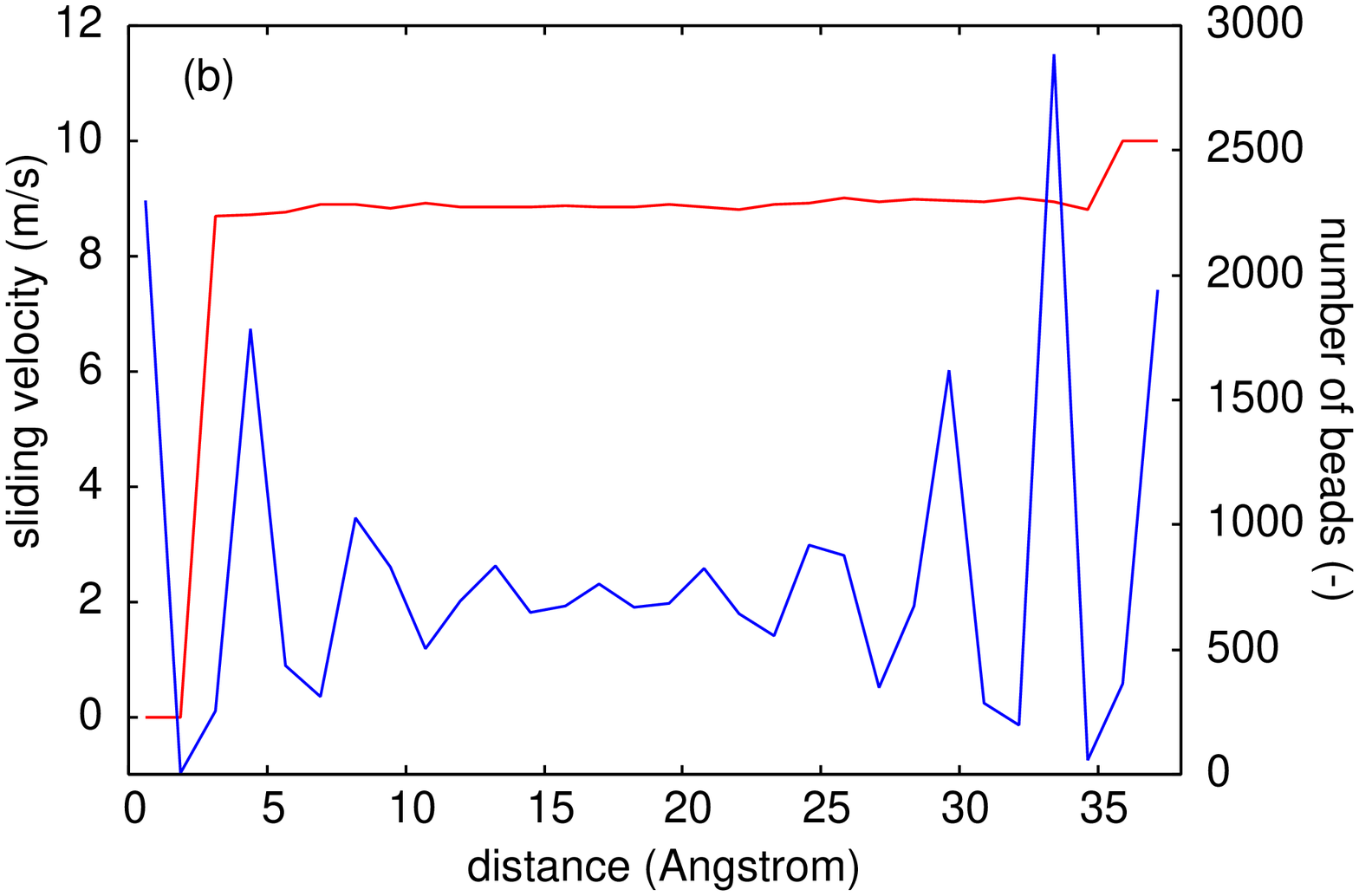}
  \label{velx_zposC100m1010_690_450_K}
  }
  \caption{
The distributions of the average number and the average velocity $v_x$ (in the sliding
direction) of lubricant
C atoms along the distance between the substrate and the block for C$_{100}$H$_{202}$
metal-polymer sliding at velocity of the block $v=10 \ {\rm m/s}$ for the normal
pressure $p=10 \ {\rm MPa}$ for the temperature a) $27^\circ {\rm C}$ and
b) $177^\circ {\rm C}$. In the latter case the slip also occurs between the block
and the last monolayer of lubricant molecules due to thermal fluctuations. The
very left and very right maximum of the both density distributions correspond to
the substrate and the block atom layers.
}
  \label{velx_zposC100m1010_300_K_450_K}
\end{figure}

For temperatures well above the melting point the molecules in
the center of the polymer film are disordered as expected for
the liquid state of the lubricant.
In Fig.~\ref{velx_zposC100m1010_300_K_450_K} we show the density distribution,
and the average velocity $v_x$ of the C-atoms along the distance between
the substrate and the block ($z$-direction) for
the C$_{100}$H$_{202}$ polymer film. The results are for $T=27^\circ {\rm C}$ and $177^\circ {\rm C}$.
At $T=27^\circ {\rm C}$ the film is seven monolayers thick,
but at $T=177^\circ{\rm C}$, due to the thermal
expansion, the system has eight monolayers. In the center the
molecules are disordered and the layers of molecules blurred.

For $27^\circ {\rm C}$ the average velocity changes abruptly at the
substrate-lubricant interface (see Fig.~\ref{velx_zposC100m1010_300_K_450_K}a), so
the slip occurs between the substrate and the first monolayer of molecules of the
polymer film. Thus, the whole polymer film is bound to the block and moves with the average
velocity $v_x \approx 10 \ {\rm m/s}$, i.e. with velocity of the block.
For $T=177^\circ {\rm C}$ most of the slip also occurs at the polymer-substrate
interface (see
Fig.~\ref{velx_zposC100m1010_300_K_450_K}b), but a small slip (slip velocity $v\approx 1 \ {\rm m/s}$)
also occurs at the polymer-block interface. Thus, all monolayers of the polymer film move
with the average velocity $v_x \approx 9 \ {\rm m/s}$.
The slip at the polymer-block interface is due to the applied shear stress and thermal fluctuations.
At the lower temperature $27^\circ {\rm C}$ the thermal fluctuations are not strong enough
to overcome the relatively large atomic corrugation at the polymer-block interface,
which results from the relatively large  $\epsilon_1 = 40 \ {\rm meV}$ for the interaction
of each polymer bead units with the block atoms ({\em strong adsorbate interaction}).
Friction is a stress-aided thermally activated process, and in the present case when the temperature increases from
$T=27^\circ{\rm C}$ to $177^\circ{\rm C}$, the shear stress drops by a factor of $\sim 2$.

The effective corrugation of the interaction potential experienced by the molecules
at the sliding interface is the most important parameter influencing the magnitude of the friction and the
dependence on the external (squeezing) pressure. Indeed, the fact that the lattice constant of the
substrate is much smaller than the size of the polymer molecules (and the polymer persistence length),
and also very different from the
natural separation between the polymer molecules, implies that the effective corrugation of the interaction
potential between the polymer and the ``metal'' substrate will be very small, and this explains
the small friction observed in this case, compared to the case when the slip occurs at the
polymer-polymer interface.

\subsection{4. Polymer on polymer}

\begin{figure}
  \includegraphics[width=0.46\textwidth]
{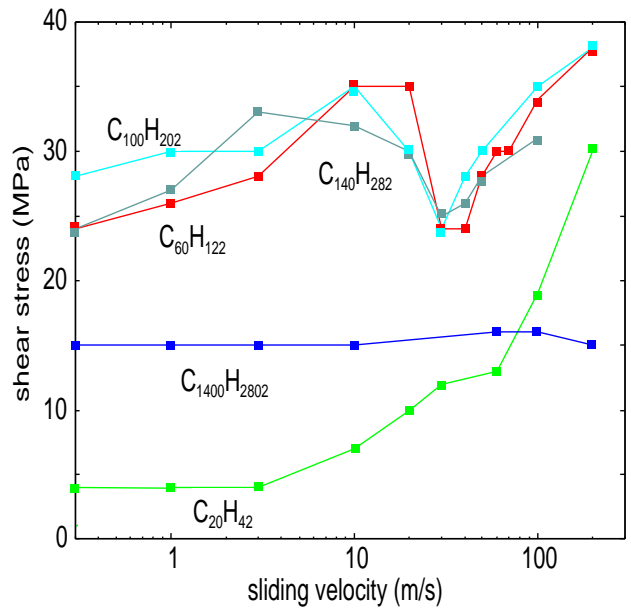}
  \caption{ \label{polpolallssV}
The shear stress as a function of the sliding velocity for all the investigated systems. The normal pressure is $10 \ {\rm MPa}$.
}
\end{figure}

When a polymeric film is strongly attached to the
block and substrate surfaces, sliding of the
block will induce a shearing of the polymer film. For this case
we have investigated the influence of the sliding speed on the
shear stress. Figure \ref{polpolallssV}
shows the results obtained for a number of linear
alkanes with chain length from 20 C-atoms to 1400 C-atoms.
Note that for C$_{1400}$H$_{2802}$ the shear stress is independent of the sliding velocity,
whereas the C$_{20}$H$_{42}$ is more liquid-like, with a shear stress approximately linearly related to the sliding
velocity.


\begin{figure}
  \includegraphics[width=0.47\textwidth]
{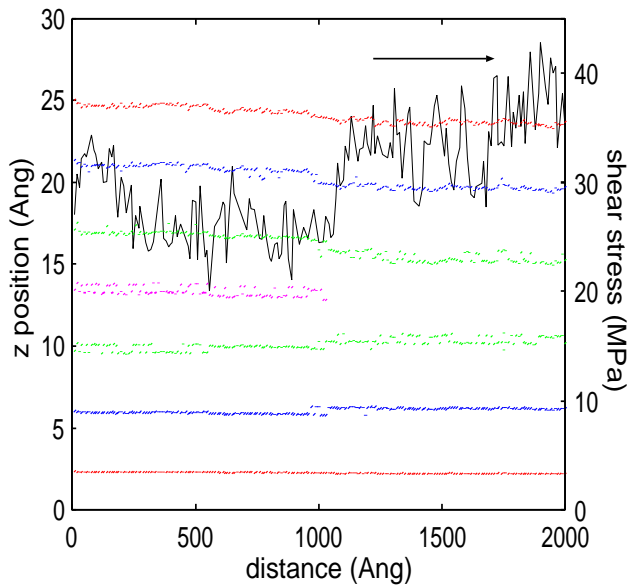}
  \caption{ \label{C100_p_10_from_100_layers}
The shear stress and the positions in the the z-direction of the layers for the C$_{100}$H$_{202}$
system as a function of the sliding distance.
 The sliding velocity of the block is changed from $100 \ {\rm m/s}$ to $10 \ {\rm m/s}$ at $0 \ {\rm \AA}$.
}
\end{figure}

For the mid-sized molecules (with 60 to 140 C-atoms in the chain), for small and large velocities the shear stress
is nearly proportional to the logarithm of the sliding velocity, with a small slope at low velocities and a larger
one at large velocities. In the range of from $20 \ {\rm m/s}$ to $40 \ {\rm m/s}$
the shear stress decreases with increasing sliding velocity.
This is due to a layering transition, where the number of layers increases from 6 to 7. This is proved
in Fig. \ref{C100_p_10_from_100_layers} which shows the positions in the the z-direction of the layers
for the C$_{100}$H$_{202}$ system as a function of the sliding distance $d$.
At $d=0$ the sliding velocity of the block is changed from
$100 \ {\rm m/s}$ to $10 \ {\rm m/s}$ and after some relaxation time period (which corresponds
to the sliding distance $d \approx 1000 \ {\rm \AA}$) the system abruptly switches from 7 to 6 layers. At the
same time the shear stress increases abruptly (see Fig. \ref{C100_p_10_from_100_layers}).
The latter is, at least in part, due to the decreased number of
slip planes. The layering transition in Fig. \ref{C100_p_10_from_100_layers} is reversible: increasing the velocity back to
$100 \ {\rm m/s}$ results in a return to 7 layers.

As pointed out above, for the mid-sized molecules, for small and large velocities the shear stress
is nearly proportional to the logarithm of the sliding velocity, with a small slope at low velocities and a larger
one at large velocities. The logarithmic velocity dependence is expected for thermally activated stress induced
processes, which predict the frictional shear stress \cite{P2}
$$\sigma_{\rm f} \sim {k_{\rm B} T \over E_{\rm B}} {\rm log} \left ({v\over v_0}\right )$$
where $k_{\rm B} T$ is the thermal energy ($T$ is the temperature),
$E_{\rm B}$ is an energy barrier (activation energy for some rearrangement process involved in
lateral slip, e.g., removal of polymer bridge) and $v_0$ is a reference
velocity. Thus, the larger slope of the
$\sigma_{\rm f}({\rm log} v)$ relation for high slip velocities, as compared
to low slip velocities, can be explained by assuming that the energy barrier $E_{\rm B}$ is
smaller in the more open structure, which prevail after the transition from 6 to 7 layers with increasing velocity.

The change in the number of layers in the film is thermally activated, since
no change in the number of layers (on the time scale of our simulations)
occurs when the thermostat is at $0 \ {\rm K}$.
Experimental data suggest that the layering transition happens at lower sliding
velocities when the normal pressure is decreased \cite {bureau2006}. This is in
good accordance with our results as a decrease in pressure from $10 \ {\rm MPa}$ to
$3 \ {\rm MPa}$ shifted the transition velocity of the C$_{60}$H$_{122}$ system
from about $30 \ {\rm m/s}$ to $20 \ {\rm m/s}$.

\begin{figure}
  \includegraphics[width=0.48\textwidth]
{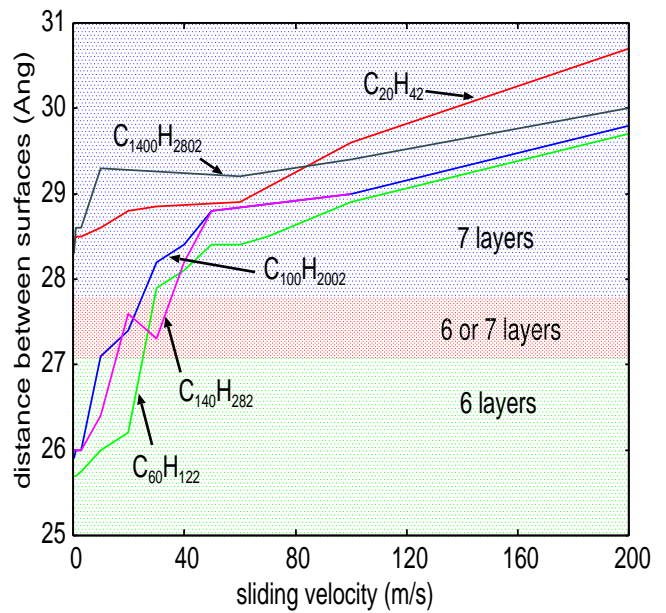}
  \caption{ \label{polpolallzV}
The distance between the surfaces as a function of the sliding velocity for all the investigated systems. There is also
an indication of the number of layers present at different surface separations. It
 was chosen to use a linear x axis to show the transition more clearly.
 The normal pressure is $10 \ {\rm MPa}$.
}
\end{figure}

We will now study the nature of the layering transition in greater detail. In Fig. \ref{polpolallzV}
we show the separation between the surfaces for each polymer system we have studied,
as a function of the sliding speed.
Note that the surface separation in the case of
C$_{20}$H$_{42}$ and C$_{1400}$H$_{2802}$ increases with increasing sliding velocity
with a constant slope, but the number of
polymer layers is constant (at seven) in these cases. For the mid-sized molecules there is an abrupt
increase in the surface separation in the transition regime of figure \ref{polpolallssV}. The number
of layers is six before the transition (below $10 \ {\rm m/s}$) and seven after (above $40 \ {\rm m/s}$).

\begin{figure}
  \includegraphics[width=0.48\textwidth]
{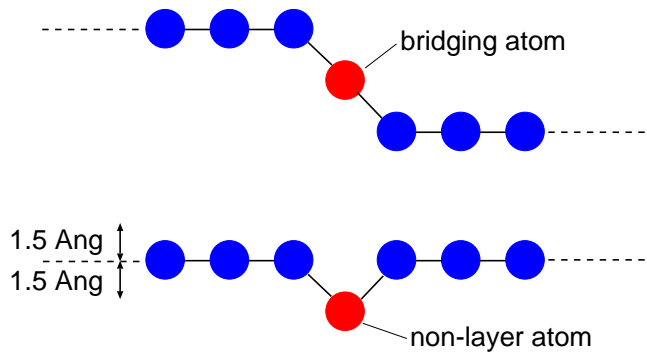}
  \caption{ \label{bridgedef}
The definitions of bridge and non-layer atoms used in the present paper.
}
\end{figure}
\begin{figure}
\includegraphics[width=0.48\textwidth,angle=0]{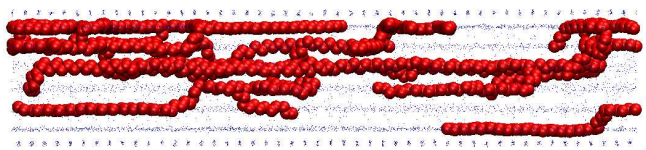}
  \caption{ \label{snap_c100_bridges}
Snapshot picture of a C$_{100}$H$_{202}$ system where ten molecules have been chosen to be shown at random. The rest of the
   atoms are reduced to points. The picture shows that the molecules have atoms in several layers. Some of these segments form bridges
   but others are just present in different layers without linking these. The normal pressure is $10 \ {\rm MPa}$ and the
   sliding velocity is $0.3 \ {\rm m/s}$.
}
\end{figure}
\begin{figure}
  \includegraphics[width=0.47\textwidth]
{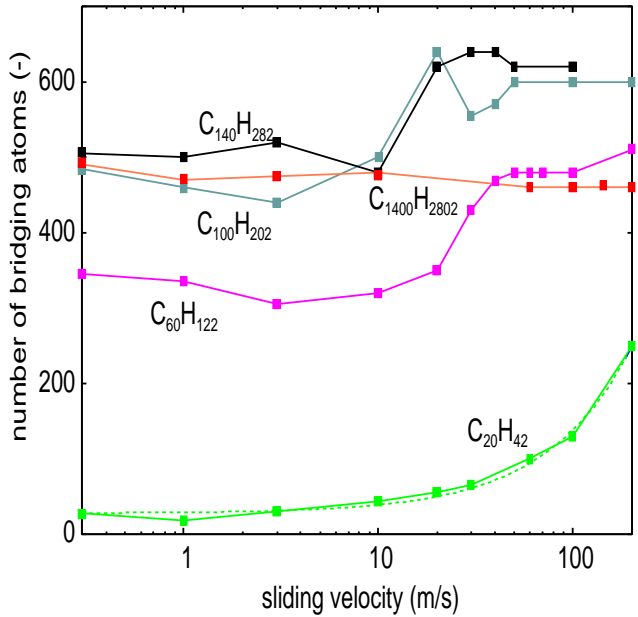}
  \caption{ \label{polpolbridges}
The number of bridging atoms as a function of the sliding velocity for the investigated systems. A bridging atom is defined in the text and in figure \ref {bridgedef}. The dotted line
is a linear fit to C$_{20}$H$_{42}$. The normal pressure is $10 \ {\rm MPa}$.
}
\end{figure}
\begin{figure}
  \includegraphics[width=0.47\textwidth]
{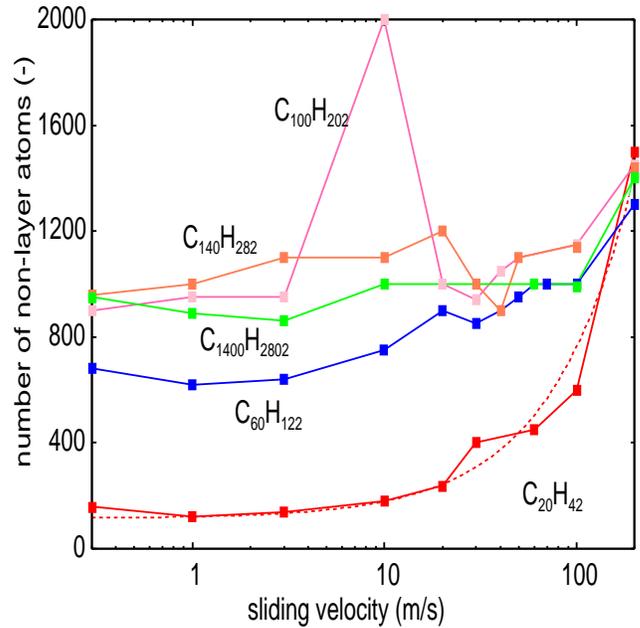}
  \caption{ \label{polpoloutside}
The number of non-layer atoms as a function of the sliding velocity for the investigated systems. A non-layer atom is defined in the text and in figure \ref {bridgedef}.
The dotted line is a linear fit to C$_{20}$H$_{42}$. The normal pressure is $10 \ {\rm MPa}$.
}
\end{figure}

Let us now study the number of molecular bridges between the layers, which may be affected by the layering transition
\cite {drummond2002,subbotin1997,bureau2006,qian2003}. There is no exact definition of a molecular bridge, but we
have chosen to define a bridging atom by the fact that it does not belong to the same layer as the
preceding atom in the molecule, and at least three of the preceding
and three of the following atoms in the molecule are in two different
layers, see Fig. \ref{bridgedef}. The bridging phenomenon can be observed in the snapshot shown in figure \ref {snap_c100_bridges}.
The number of bridging atoms as a function of the sliding velocity is shown in figure \ref{polpolbridges}.
Note that the C$_{1400}$H$_{2802}$ system has the same number of bridges through the whole
velocity range. The C$_{20}$H$_{42}$ system shows a linear dependence, whereas the mid-sized molecules
seem to have a constant number before the transition regime, increasing to a higher level after this regime.
The latter can be explained by the more open structure of the 7-layer systems, which makes it easier for polymer
molecules from one layer to have segments extending into other layers.
Figure \ref{snap_c100_bridges} also shows that some atoms are outside the center line of the polymer
layers without linking these. We define such a non-layer atom as one having a distance of
at least $1.5 \ {\rm \AA}$ to the center line of any layer (see Fig. \ref{bridgedef}).
We present the number of non-layer atoms as a function of the sliding velocity in figure \ref{polpoloutside}.

Figure \ref {polpoloutside} shows that the C$_{1400}$H$_{2802}$ system has an increasing number
of non-layer atoms with increasing sliding velocity. In the case of C$_{20}$H$_{42}$ the number of non-layer
atoms is proportional to the sliding velocity. The increase in the number of non-layer atoms is associated
with the increase in the separation between the layers with increasing sliding velocity (see Fig. \ref{polpolallzV}),
which allow polymer segments in a layer to displace away from
the layer-plane by a considerable distance without experience a large repulsion from the nearby polymer layer.
One may alternatively interpret the increase in the layer separation as resulting from the increased repulsion from
the non-layer atoms with increasing sliding velocity. In this picture the increased number of non-layer atoms results from
the increased momentum transfer in collisions between atoms in two nearby layers as the sliding velocity increases: these
collisions kick polymer segments away from the layer-plane.

The C$_{60}$H$_{122}$, C$_{100}$H$_{202}$ and the C$_{140}$H$_{282}$ systems exhibit a slow increase in
the number of non-layer atoms before
the transition, and a faster one when the polymer film has increased to seven layers.
Note that the C$_{100}$H$_{202}$ system exhibits a maximum in the number of non-layer atoms around $10 \ {\rm m/s}$.
This is caused by a strong fluctuation which sometimes was observed
also for the 60 C-atom and 140 C-atom
systems (not shown). This is not unexpected as strong fluctuations often occur close to phase transition
points (in this case a layering transition).




\begin{figure}
  \includegraphics[width=0.5\textwidth]
{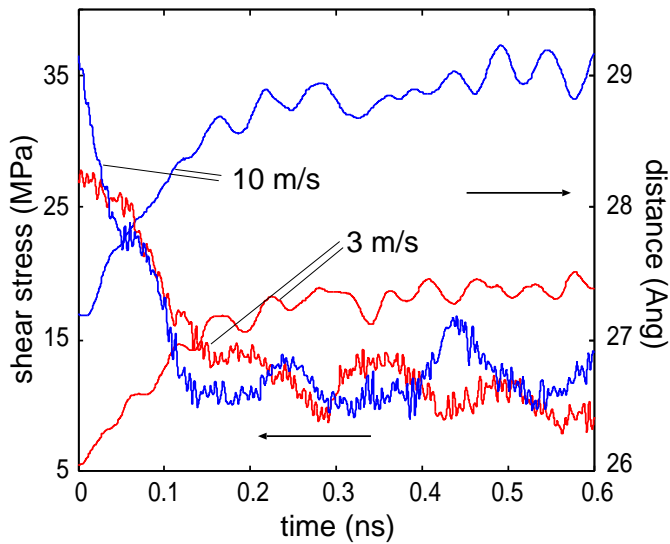}
  \caption{ \label{C100_p_10_3_10_melt}
The shear stress and the distance between the surfaces as a function of the sliding time when the
temperature is increased from 300 K to 450 K. The melting point is 390 K. The sliding velocities are
3 and 10 m/s.  The normal pressure is $10 \ {\rm MPa}$. During melting the film increases its
number of layers from 6 to 7 at 10 m/s whereas it stays at 6 layers at 3 m/s. The system is C$_{100}$H$_{202}$.
}
\end{figure}

Let us now compare the transition observed in figure \ref{polpolallssV} with
what happens during melting of the polymer film. We have investigated the behavior of the film when
it is melted by raising the temperature from 300 K to 450 K;  see figure \ref{C100_p_10_3_10_melt}.
The figure shows the shear stress and the distance between the surfaces as a function of the sliding time.
For both 3 and 10 m/s the shear stress decreases to about $10 \ {\rm MPa}$, a much lower level than observed
in figure \ref{polpolallssV}. At 3 m/s the film keeps its 6 layers as the melting increases the distance
between the surfaces to about $27.4 \ {\rm \AA}$, a result that is consistent with figure \ref{polpolallzV}.
The combination of a higher sliding velocity and melting increases the distance between the surfaces to
about $29 \ {\rm \AA}$ when the sliding velocity is 10 m/s. The film then passes to 7 layers as predicted
by figure \ref{polpolallzV}.  From this we deduce
that the layering transition induced by the increase in the sliding velocity is not associated with
a melting of the film. The transition seems to occur at a certain surface separation and the shear stress
is not affected by it in the melted state whereas this is the case when the sliding velocity is increased.

\begin{figure}
  \includegraphics[width=0.46\textwidth]
{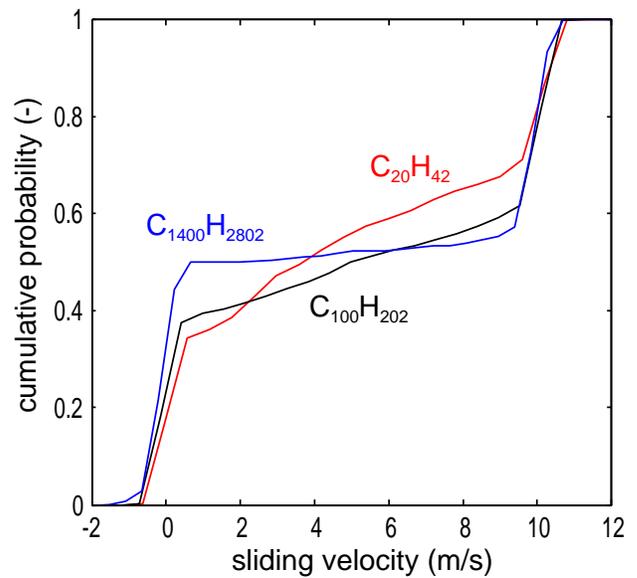}
  \caption{ \label{polpolaccnV}
 Sliding velocity of atoms  as a function of the number of  atoms sliding at
 less than this specific sliding velocity. The number of atoms is accumulated so that the total
 reach 20000 for C$_{20}$H$_{42}$ and C$_{100}$H$_{202}$ whereas the C$_{1400}$H$_{2802}$
 system only has 19600 atoms.
The sliding velocity of the block is $10 \ {\rm m/s}$ and the
 normal pressure is $10 \ {\rm MPa}$.
}
\end{figure}

Let us study the velocity profiles of the systems.
Figure \ref {polpolaccnV} shows the cumulative velocity probability distribution.
Note that for the C$_{1400}$H$_{2802}$ system nearly all the slip occurs at one interface at the center of the polymer film.
Thus, this system can be considered as two polymer slabs sliding against each other. This picture also explains the
independence of the shear stress for C$_{1400}$H$_{2802}$ to the number of bridges
(figure \ref {polpolbridges}) and non-layer atoms (figure \ref {polpoloutside}), since these are presumably
mostly appearing inside the polymer slabs, and thus do not influence the
interfacial slip. The C$_{20}$H$_{22}$ molecules have their sliding planes
distributed over the whole thickness of the film. This is expected for a liquid-like
flow, but figures \ref {polpolbridges} and \ref {polpoloutside} indicate that this also could be a result of
an increasing number of bridges and/or non-layer atoms. The mid-sized molecules in
figure \ref {polpolaccnV} have a velocity profile in between the longest molecules
and the shortest ones, and can be considered as a so-called plug flow,
where the outermost layers are pinned to the surfaces of the block and the substrate.

\begin{figure}
  \includegraphics[width=0.48\textwidth]
{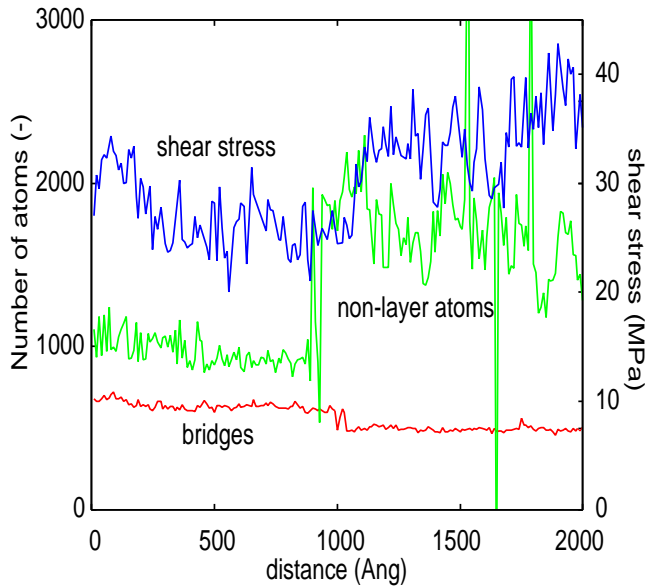}
  \caption{ \label{C100_p_10_from_100}
The number of bridging and non-layer atoms and the shear stress as a function of the sliding distance for C$_{100}$H$_{202}$.
 The sliding velocity of the block is changed from $100 \ {\rm m/s}$ to $10 \ {\rm m/s}$ at $0 \ {\rm \AA}$.
}
\end{figure}

In Fig. \ref{C100_p_10_from_100}, at $d= 0 \ {\rm \AA}$ (where $d$ is the slip distance)
the sliding velocity of the block is changed from $100 \ {\rm m/s}$ to $10 \ {\rm m/s}$.
Note that as the film relaxes the number of bridges is quite constant
whereas the number of non-layer atoms is decreasing slightly and so is the shear stress.
When the transition regime is reached the number of non-layer atoms suddenly increases
as the central layers become mobile and (in the $z$-direction) diffuse.
The increase in shear stress follows this increase
in the number of non-layer atoms whereas the number of bridges decreases to a lower constant level as
the number of layers goes from seven to six. This can be understood as the density of
molecules in the layers is smaller for the 7-layer state than for the 6-layer state,
and hence the ability for chain molecules to (due to a fluctuation) interdiffuse
and form bridges will be largest in the 7-layer state.

The shift from six to seven layers increases the number of slip planes by one.
At the same time the density of bead-units in the layers decreases
so that the space between the bead units in each layer increases.
The increased space reduces the energetic barriers for polymer segment rearrangement
processes, and results in the formation of more cross-links (see
figure \ref {polpoloutside}). At low velocities the shear stress increases
slowly with the increase of the sliding velocity, but after the
transition the slope becomes steeper (see figure \ref{polpolallssV}).
This can again be attributed to reduced energetic barriers for polymer rearrangement processes.

\subsection{5. Summary and conclusions}

We have presented results of molecular dynamics calculations of friction
for two solids separated by
a $\approx 3 \ {\rm nm}$ thick polymer film.
Two types of systems were considered: (a) a polymer film pinned to one
of the solid surfaces and sliding at the other solid surface (the
``metal''-polymer case). The second case (b) was with the polymer layers pinned to both solid
surfaces and shearing at the polymer-polymer interface (the polymer-polymer case).
We used linear alkane molecules with the number of carbon atoms 20, 60, 100, 140
and 1400.

The frictional shear stress for the polymer-polymer systems is much higher than for
the ``metal''-polymer systems. This is due to the same size of the atoms or molecules on
both sides of the slip-plane for the polymer-polymer case, resulting in strong interlocking (as
for a commensurate interface), while the ``metal''-polymer interfaces are incommensurate
(the lattice constant of the ``metal'' substrate is different from the distance between
atoms of the lubricant molecules).

We have studied the
velocity dependence of the frictional shear stress for both cases. In the first setup
the shear stresses are relatively independent of molecular length.
For the shortest hydrocarbon C$_{20}$H$_{42}$ the frictional shear stress is lower
and increases approximately linearly with the velocity.

For polymer sliding on polymer
the friction is significantly larger, and the velocity dependence is more complex. For
the longest molecules (1400 carbon atoms) the shear stress is independent of the sliding
velocity as the sliding occurs primarily at one interfacial slip plane. The shortest molecules
again exhibit liquid-like sliding with the shear stress being
approximately proportional to the sliding velocity.
The mid-sized molecules (60 to 140 C-atoms) show a slightly increasing shear stress at low velocities,
and a faster increase at high sliding velocities. Between these regimes there is a
transition with a decrease  in the shear stress with increasing sliding velocity.

The mechanism behind this behavior seams to be a kinetic phase transition involving
a change in the number of layers
in the film, which introduce new slip planes. This decreases the shear
stress abruptly. Further increase of the sliding velocity will increase the shear
stress rapidly, which we attribute to the interaction between the layers via non-layer atoms.


\subsection{Acknowledgments}

A part of the present work was carried out in frames of the European Science Foundation
EUROCORES Programme FANAS supported from the EC Sixth Framework Programme,
under contract N. ERAS-CT-2003-980409.
Two of the authors (I.M.S. and V.N.S.) acknowledge support from
IFF, FZ-J\"ulich, hospitality and help of the staff during their research visits.

\end{document}